# COVID-19 and Social Distancing: Disparities in Mobility Adaptation between Income Groups

Kentaro Iio[1*], Xiaoyu Guo[2], Xiaoqiang "Jack" Kong[2], Kelly Rees and Xiubin Bruce Wang[2]

[1]*Traf-IQ, Inc., Houston, Texas 77079, United States*

[2]*Zachry Department of Civil & Environmental Engineering, Texas A&M University, College Station, Texas 77843-3136, United States*

(Updated March 12, 2021)

In response to the coronavirus disease 2019 (COVID-19) pandemic, governments have encouraged and ordered citizens to practice social distancing, particularly by working and studying at home. Intuitively, only a subset of people have the ability to practice remote work. However, there has been little research on the disparity of mobility adaptation across different income groups in US cities during the pandemic. The authors worked to fill this gap by quantifying the impacts of the pandemic on human mobility by income in Greater Houston, Texas. We determined human mobility using pseudonymized, spatially disaggregated cell phone location data. A longitudinal study across estimated income groups was conducted by measuring the total travel distance, radius of gyration, number of visited locations, and per-trip distance in April 2020 compared to the data in a baseline. An apparent disparity in mobility was found across estimated income groups. In particular, there was a strong negative correlation ($\rho$ = -0.90) between a traveler's estimated income and travel distance in April. Disparities in mobility adaptability were further shown since those in higher income brackets experienced larger percentage drops in the radius of gyration and the number of distinct visited locations than did those in lower income brackets. The findings of this study suggest a need to understand the reasons behind the mobility inflexibility among low-income populations during the pandemic. The study illuminates an equity issue which may be of interest to policy makers and researchers alike in the wake of an epidemic.

---

## HIGHLIGHTS

- We investigated mobility in Greater Houston by income during the COVID-19 pandemic.
- Spatially disaggregated mobility data in April 2020 and a baseline were analyzed.
- Higher income was associated with larger mobility reduction in multiple measures.
- A need for addressing inequity in urban mobility during the epidemic was apparent.

## 1. INTRODUCTION

THE pandemic of coronavirus disease 2019 (COVID-19) has had a significant worldwide impact in terms of health and economy (Acter et al., 2020; Torales et al., 2020). As of January 30, 2021, the total number of confirmed cases has surpassed 100 million globally and 25 million in the United States alone. The death toll by this same date reached over 2.1 million globally and 429,000 in the United States (World Health Organization, 2021).

In the absence of effective pharmaceutical treatments, one of the practices believed to be effective at limiting the spread of COVID-19 was reducing the degree of physical contact with others and minimizing exposure to the virus. Governments have aimed to "flatten the curve" of infections by introducing measures to decrease physical contact between individuals, including implementing travel restrictions, enforcing border closures, and encouraging social distancing (Block et al., 2020; Chinazzi et al., 2020; de Haas et al., 2020; Hsiang et al., 2020; Phan and Narayan, 2020; Ruktanonchai et al., 2020). In 2020, the United States federal government implemented travel restrictions and travel warnings for heavily-infected regions beginning in late January and early February, and state and local governments issued "stay-at-home" and social distancing orders by mid-March (Peirlinck et al., 2020). As a result, American travel behaviors have dramatically changed in terms of origins, destinations, modes, and travel frequency due to COVID-19 and the associated health and government mobility orders (Barbieri et al., 2020; De Vos, 2020).

* Corresponding author (kiio@traf-iq.com)

While these temporary restrictions on human mobility have negatively affected short-term economic and employment growth (Nicola et al., 2020), they showed some positive disease-containment effects in May and June 2020; during these two months, newly confirmed COVID-19 cases and fatalities dropped in many regions (Block et al., 2020; Courtemanche et al., 2020; Thu et al., 2020). Improvements were assumed to be at least partially due to the public's observance of the executive orders. However, different groups of the population may have engaged in different degrees of change in behaviors as shown by travel distance, frequency, modal shift, work schedules, and remote working (Kramer and Kramer, 2020). In fact, medical records suggest that individuals within urban areas have faced different hospitalization rates and fatality rates per capita depending on their economic status (Raifman and Raifman, 2020). In addition, a longitudinal survey in the United States indicates a smaller proportion of lower-income respondents shifted to remote work by July than did higher-income respondents (Circella, 2020). The different changes in behavior within different groups are worthy of a special study in order for policy makers to make effective policies and maintain social equity in future decision making.

## 2. LITERATURE REVIEW

The literature reviewed for this study is organized according to several areas as follows.

### 2.1 Human mobility and location characteristics

As mobile communication devices have integrated into society over the past two decades, researchers have gained insights through the location data provided by these devices. Nowadays, it has been known that travelers' mobility is well characterized by variables such as the travel distance, approximate spatial range of travel, and the frequency of uniquely visited locations. González et al. (2008) defined an approximate range of activity space as the "radius of gyration" (the average distance to observed locations from the center of mass of all sets of observations for an individual traveler). Based on estimated mobile phone user trajectories within Voronoi cells, the researchers found that the trip distance distributions can be modeled with great accuracy as a function of the radius of gyration, which is a function of the measurement period. Additional literature reports that individual human mobility follows highly repetitive, predictable patterns (e.g., home-office and home-school) on a daily basis (Schneider et al., 2013; Song et al., 2010). Therefore, it is possible to estimate one's socioeconomic variables based on data obtained from mobile devices (Frias-Martinez et al., 2010; Soto et al., 2011), though the estimation accuracy depends on urban structures. As Prestby et al. (2019) found in Milwaukee, Wisconsin, individual human mobility patterns seem to have clear differences among different socioeconomic groups especially where practical residential segregation exists.

### 2.2 COVID-19 and travelers' mobility

In the hope of revealing the effects of mobility change on the spread of this infectious disease, scholars have investigated cell phone location data to report overall reductions in human mobility measures during the COVID-19 pandemic. In Italy, Pepe et al. (2020) analyzed anonymized cell phone location data from Cuebiq, Inc. aggregated by province and found a reduction greater than 30 percent in total trips and in the radius of gyration during the country's first lockdown compared to baselines before the COVID-19 outbreak. Bonaccorsi et al. (2020) investigated proprietary mobility data from Facebook, Inc. aggregated by Italian municipalities and reported larger mobility reductions in municipalities with higher fiscal capacity.

In Tokyo, Japan, Yabe et al. (2020) analyzed cell phone location data provided by Yahoo! Japan Corporation and reported a negative correlation between assumed taxable income per household and the reduction in mobility during the country's state of emergency. They also mention the scarcity of research on how human mobility, stratified by different economic statuses, has been affected by the pandemic.

Multiple mobility studies have been conducted in the United States. Weill et al. (2020) performed a panel regression analysis on aggregated mobility metrics by census tract and county across the United States. Based on four mobility metrics (i.e., "Completely at Home" (SafeGraph, Inc.), Median Distance Traveled (SafeGraph, Inc.), Device Exposure (PlaceIQ, Inc.), and "Retail and Recreation" (Google, LLC)) by aggregated income quantiles, they concluded high-income areas had larger reductions in mobility than low-income areas. Their model considered the effect of federal policies which applied to all counties. Jay et al. (2020) compared SafeGraph mobility metrics across US census block groups by aggregated income quintile. Using January and February 2020 as the baseline, a difference-in-differences linear regression revealed that in April, people in high-income neighborhoods spent more time at home compared to those who lived in low-income neighborhoods. At a regional level, Apple, Inc. (2020) and Google, LLC (2020) reported an approximate 60 percent reduction in the number of routing requests within map services in Houston, Texas, in April 2020 from their assumed baselines.

## 2.3 Research gaps and contribution

Most existing human mobility research on COVID-19 uses nationwide aggregated zonal mobility statistics at either state or county levels. As of early 2021, few longitudinal studies have been conducted on COVID-19-related mobility using spatially disaggregated cell phone location data. It is hard to conclude at a nationwide level that the observed differences in mobility patterns were due to inequity rather than to actions based on enacted policies or personal beliefs. In fact, observed mobility patterns also varied across states that imposed different degrees of travel restrictions (Abouk and Heydari, 2021). The linkage between human mobility patterns during the pandemic and economic status are especially worth studying within a region under a uniform policy. It also should be noted that most existing research in this topic relies on proprietary mobility metrics, so fundamental variables characterizing human mobility (e.g., radius of gyration) have rarely been reported. If records correctly suggest that the disproportionality in COVID-19 fatalities is correlated with disparity among income groups of the population and their different levels of mobility, it would naturally allude to a possible correlation between economic status and change in mobility during the pandemic.

This effort aims to fill the research gap by investigating the impact of COVID-19 and the associated containment policies on the mobility of individual travelers with varying income levels through spatially disaggregated mobility data from one metropolitan area: Greater Houston (Houston-The Woodlands-Sugar Land). The authors approach this objective through observing multiple mobility measures (i.e., total trip distance, radius of gyration, number of distinct visited locations, and per-trip distance) in April 2020, during which an executive order to restrict "non-essential" travels was in effect throughout Texas.

To the authors' knowledge, this study appears to be one of the first longitudinal studies using spatially disaggregated travel data to reveal mobility change from the income perspective in the United States. Results regarding mobility changes across income brackets would allow policy makers to make informed decisions regarding economic viability and social equity.

Paralleling existing literature, our hypothesis is that higher income is associated with larger absolute and relative mobility reductions in April 2020 compared to the baseline of January and February 2020.

The following subsection includes a brief description of the Texas COVID-19 timeline through April 2020, with a focus on the Greater Houston area, to provide context for the analysis presented in this study. Readers familiar with the COVID-19 situation in Texas may skip to the following section that describes the methodology.

## 2.4 COVID-19 timeline in Texas

The first confirmed-positive case of COVID-19 in Texas was reported in Fort Bend County on March 4, 2020, with about a dozen more cases quickly following (Texas Department of State Health Services, 2020a). The first confirmed death caused by COVID-19 in Texas was reported in Matagorda County on March 14 (Texas Department of State Health Services, 2020b). The Governor of Texas declared a state of disaster on March 13 for all counties in Texas (Office of the Texas Governor, 2020). Over the next few days, metropolitan areas including Houston ordered the closure of restaurants and bars and imposed stricter guidelines on the size of social gatherings.

The governor instituted GA-14 at the end of March which minimized social gathering and encouraged social distancing from April 2 through April 30, 2020 (Abbott, 2020). Additionally, the Harris County judge implemented a "Stay Home, Stay Safe" order from March 24 through April 24, making it one of 51 counties in Texas to enact "stay-at-home" orders by the end of March (Harris County, 2020; Wilson, 2020). As of April 30, Texas had over 28,000 confirmed cases and 780 confirmed fatalities (Centers for Disease Control and Prevention, 2020; Johns Hopkins University & Medicine Coronavirus Research Center, 2020).

The effects of the stay-at-home orders on human mobility were apparent in Harris County. According to the Bureau of Transportation Statistics (BTS), the percentage of people not traveling from home in Harris County, Texas, was reported as approximately 16.9 percent, or approximately 794,800 people, in January (Bureau of Transportation Statistics, 2020); these people are assumed to stay home for reasons other than the global COVID-19 pandemic. In March and April, this percentage increased to 20 percent (945,300 people), and 23.6 percent (1,107,800 people), respectively; in May and June, it dropped back toward the baseline with 19.3 percent (906,600 people) and 19.7 percent (925,200 people) respectively. Within Harris County, BTS used anonymized mobile phone data to record the average number of daily trips as approximately 15,952,000 and 16,736,000 trips in January and February. As people began to work from home, the average number of daily trips dropped to about 14,593,000 and 12,028,000 trips in March and April, respectively

## 3. METHODS

The authors used Microsoft Office 365 ProPlus, PostGIS 3.0, QGIS 3.14, and Julia 1.4.1 (Bezanson et al., 2017) to investigate spatiotemporal changes in human mobility through smartphone location data along with projected income.

### 3.1 Mobility data and samples

Pseudonymized iOS mobility data in January, February, and April 2020 were provided by SAFE2SAVE, LLC, a Texas-based company operating a smartphone application (app). The data contained 89,928,723 data points on roadways, and each data point contained a pseudonymized user identification number (random integers), geographical coordinates, and a timestamp. The data contain no personal information such as name, income, or age.

In order to examine the changes in mobility patterns during the global COVID-19 pandemic, mobility data in April 2020 were compared to the average data in January and February 2020 (baseline). April 2020 included the entire duration of GA-14 (0:00:01 a.m. on April 2, 2020, through 11:59:59 p.m. on April 30, 2020), the executive order which encouraged residents to stay home except for essential travels. March 2020 was not included in the study due to the varying implementation dates of stay-at-home orders in different jurisdictions.

To ensure the April data points were comparable to the baseline, analyses included only data of those who recorded (i) at least one data point in April and (ii) at least one data point per week over the eight consecutive weeks between January 5 and February 29. The second filtering intended to filter out people who quit using the app or non-regular app users since comparisons cannot be drawn over time when one deletes the app during the study period. Unlike spatially aggregated data, this is an apparent tradeoff a spatially disaggregated longitudinal study inherits. Although this sample reduction might have affected the results, the effect was assumed to be negligible or rather suitable than the do-nothing alternative with respect to the study objective.

After this procedure, 46,047,382 data points from 26,059 people were kept for analysis. These data points correspond to approximately 6.6 million person trips.

### 3.2 Variables

As an independent variable, per-capita income in the 2018 American Community Survey (ACS) (United States Census Bureau, 2020) was used. In the United States, one census tract represents a community that represents a somewhat-homogeneous cluster of approximately 4,000 residents. The ACS was overlaid at the resolution of United States census tracts using a circular area with 0.5-mi diameter around each traveler's median latitude and longitude of the first and last records of each day during the study periods. When the circular cordon overlapped multiple census tracts, the weighted average of the census incomes was considered to be that person's most probable income. While each observation was collected on a roadway, it is reasonable to assume each observation was geographically in or near users' residential neighborhood when the data were aggregated (Chen et al., 2014). This geographical assumption is a methodological limitation of our research. However, a reasonable level of accuracy was expected because Houston is known as one of the most residentially segregated metropolises in the United States (Fry and Taylor, 2012). Overlaying income typically commits an ecological fallacy to some extent; however, the segregated nature of Houston neighborhoods lessens this fallacy, and the projected income in this paper are conceptually similar to the likely degree of financial appanage. The procedure narrowed down our samples to 10,398 individuals who were presumed to reside within Greater Houston (Fig. 1).

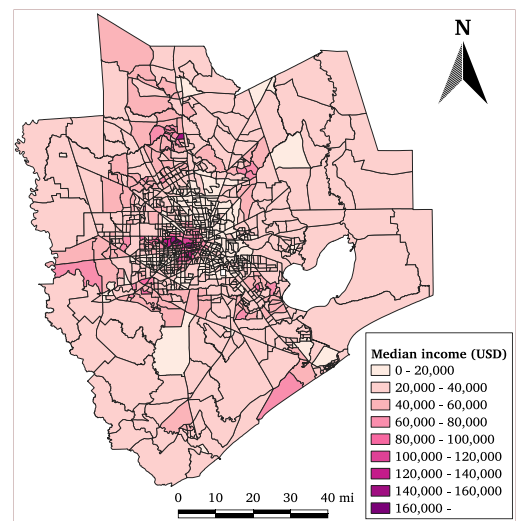

Fig. 1. Per-capita income by census tract in the Greater Houston area.

*Note.* White areas indicate the lack of income data.

The authors were interested in how long, how often, and within what distance range people traveled during the COVID-19 pandemic compared to the baseline. To answer these questions, the monthly total travel distance, radius of gyration, number of distinct visited locations, and per-trip distance were computed.

### 3.2.1 Total travel distance

Monthly total travel distances were calculated as the sum of Euclidean travel distances in a series of geographical coordinates recorded every 1.84 minutes on average. Because map matching was not used, the computed values are expected to be smaller than travel distances on roadways. Nevertheless, this calculation (Equation 1) was considered a reasonable means to observe mobility changes within subjects.

$$L^a(t) = \sum_{i=2}^{n_c^a} d_{\vec{r}_{i-1}^a, \vec{r}_i^a} \qquad (1)$$

where:  $t$ = time period

$L^a(t)$ = total travel distance of user $a$ in time period $t$

$d_{j,\ k}$ = Euclidean travel distances between the $j$th position and $k$th position

$\vec{r}_i^a$ = the $i$th position recorded for user $a$

$n_c^a$ = the number of positions recorded for user $a$

### 3.2.2 Radius of gyration

The radii of gyration for each user were calculated as follows (Equations 2 and 3):

$$\vec{r}_{cm}^a = \frac{1}{n_c^a(t)} \sum_{i=1}^{n_c^a} \vec{r}_i^a \qquad (2)$$

$$r_g^a(t) = \sqrt{\frac{1}{n_c^a(t)} \sum_{i=1}^{n_c^a} \left(\vec{r}_i^a - \vec{r}_{cm}^a\right)^2} \qquad (3)$$

where:  $\vec{r}_{cm}^a$ = the center of mass of the trajectory geographic coordinates

$r_g^a(t)$ = radius of gyration as a function of time period $t$

$P^a(t)$ = per-trip distance of user $a$ in time period $t$

### 3.2.3 Number of distinct visited locations

The temporally closest records of more than 400 seconds apart were considered as destinations and origins of consecutive trips.

All locations were aggregated with a regular hexagonal mesh of 3 km² (1.16 mi²) (c.f. Song et al. (2010)). The number of distinct visited locations ($S$) was calculated using the first and last recorded locations of each trip. This operation resulted in the averages of 3.39 trips per day per person in January and 3.54 trips per day per person in February, which are close to the statewide estimates by the Bureau of Transportation Statistics (2020). Overlapping visits were not counted even when one mesh had recorded multiple visits by one user.

### 3.2.4 Per-trip distance

Per-trip distances were derived as the arithmetic mean of the travel distance of trips made by a user (Equation 4).

$$P^a(t) = \frac{1}{m_c^a(t)} \sum_{i=2}^{n_c^a} d_{\vec{r}_{i-1}^a, \vec{r}_i^a} \qquad (4)$$

where:  $P^a(t)$ = per-trip distance of user $a$ in time period $t$

$m_c^a$ = the number of trips recorded for user $a$

Fig. 2 illustrates the flow of computing variables. All dependent variables were computed on a monthly basis ($t$ = one month).

### 3.3 Statistical Analysis

We hypothesized that there would be negative correlations between travelers' mobility measures and estimated economic status. Because the income variable was an estimate, the individuals were grouped by projected per-capita income bracket with a 10,000-dollar interval. Income brackets that had fewer than 100 people were included with the nearest bracket. This approach resulted in a smaller class interval than that of Jay et al. (2020) or Weill et al. (2020).

The distribution of trip distance, or the total distance traveled over the duration of a trip, is known to have a long tail, meaning a small percentage of travelers disproportionately influence the mean and standard deviations of aggregated mobility variables (González et al., 2008). Because the analysis of interest was mobility trends by income groups, each dataset's median was used as average for further comparisons. Spearman's rank correlation was used on medians to examine the hypothesis because the linearity of the variables over income was not necessarily assumed.

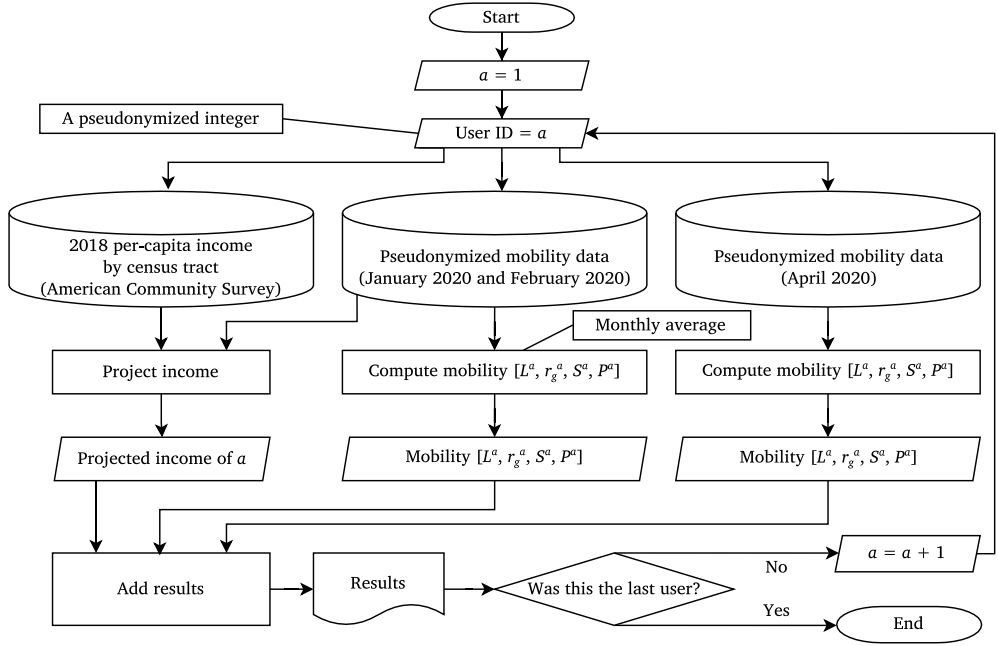

Fig. 2. Flowchart of data processing.

## 4. FINDINGS

Descriptive statistics of total travel distance, radius of gyration, and number of distinct visited locations are summarized in Table 1. The data revealed April experienced 54.7 percent, 73.6 percent, 48.1 percent, and 4.3 percent reductions in mean total trip distances, mean radius of gyration, mean number of distinct visited locations, and per-trip length, respectively, from the baselines. The similar overall percent reduction in the total travel distances (61.46 percent) was consistent with Apple, Inc. (2020) and Google, LLC (2020) as they report an approximately 60 percent reduction in the number of routing requests in April compared to January.

Fig. 3 shows median total travel distance, radius of gyration, number of distinct visited locations, and per-trip distance by projected per-capita income bracket.

Except for the per-trip distance, the values were lower in April than in the baselines, indicating the overall reduction in mobility. Because this study does not necessarily assume a linear relationship between per-capita income and the mobility variables, the Spearman's rank correlation coefficient, $\rho$, was used to examine the rank correlations among the income group. The total travel distance had almost no correlation ($\rho$ = -0.05) with the projected income bracket in the baseline, but it showed a strong negative correlation ($\rho$ = -0.90) with the income bracket in April. The estimated income bracket had a strong positive correlation with the radius of gyration in the baseline condition

($\rho$ = 0.93). However, the correlation turned out to be strongly negative in April ($\rho$ = -0.83). The number of distinct visited locations had strong negative correlations with the income level both in the baseline ($\rho$ = -0.97) and in April ($\rho$ = -0.98). Unlike the other variables, median mean per-trip distance was increased in April. While per-trip distance had a strong negative correlation with the income bracket in the baseline ($\rho$ = -0.90) and in April ($\rho$ = -0.76), the differences did not seem as evident as the other variables.

Fig. 4- 7 present cumulative distribution functions of total travel distance, radius of gyration, number of distinct visited locations, and per-trip distance by projected income bracket, respectively. It is evident that people with higher projected income brackets had larger mobility reductions in total travel distance, radius of gyration, and number of distinct visited locations. For instance, median total travel distance was reduced by 81.0 percent for individuals with a projected income group of $80,000 or larger, but it was reduced by only 62.0 percent for individuals with a projected income bracket of $20,000 or less, compared to baseline travel distances. There was a strong negative correlation ($\rho$ = -0.90) between income and total travel distance when comparing April values to baseline values. There was a strong negative correlation ($\rho$ = -0.93) between income and radius of gyration when comparing April values to baseline values. Median radius of gyration was reduced by 72.4 percent for individuals with a projected income bracket of $80,000 or larger, but it was reduced by only 32.0 percent for individuals

with a projected income bracket of $20,000 or less, compared to the baseline. Above the 70th percentile, all income groups showed somewhat similar radii of gyration during April, meaning their restricted activities happened within similar sizes of areas while they all observed the stay-at-home and social distancing orders. There was a strong negative correlation ($\rho$ = -0.86) between income and number of distinct visited locations when comparing April values to baseline values. The number of distinct visited locations was reduced by 59.6 percent for individuals with a projected income bracket of $80,000 or larger, but it was reduced by only 47.5 percent for individuals with a projected income bracket of $20,000 or less, compared to baseline travel distances. The figure also suggests smaller changes in mobility among different economic brackets past $50,000 a year, as supported by all the three aforementioned measures. In median per-trip distance, there was a moderate negative rank correlation between per-capita income and the change in April from the baseline ($\rho$ = -0.48). The drop in the radius of gyration for the higher-income groups can be partially explained in the context of larger baseline radii of gyrations in baselines (January and February).

Table 1. Descriptive statistics of dependent variables

| | Sample size | Total travel length (mi) | | Radius of gyration (mi) | | Number of visited locations | | Per-trip length (mi) | |
|---|---|---|---|---|---|---|---|---|---|
| | $n$ | $M$ | $SD$ | $M$ | $SD$ | $M$ | $SD$ | $M$ | $SD$ |
| $0 - $20,000 | 895 | | | | | | | | |
| Baseline | | 1,224.64 | 923.11 | 27.77 | 105.18 | 65.15 | 33.65 | 6.68 | 8.90 |
| April | | 624.82 | 768.84 | 9.25 | 24.44 | 38.59 | 33.94 | 6.70 | 4.85 |
| $20,000 - $30,000 | 2,525 | | | | | | | | |
| Baseline | | 1,199.49 | 879.80 | 28.31 | 76.02 | 61.49 | 33.07 | 7.20 | 10.00 |
| April | | 553.89 | 645.34 | 10.57 | 32.47 | 34.35 | 29.71 | 6.84 | 5.08 |
| $30,000 - $40,000 | 2,953 | | | | | | | | |
| Baseline | | 1,199.08 | 948.64 | 38.03 | 129.51 | 55.83 | 28.72 | 7.35 | 10.54 |
| April | | 477.13 | 607.56 | 10.10 | 27.84 | 29.33 | 27.72 | 6.97 | 5.96 |
| $40,000 - $50,000 | 2,290 | | | | | | | | |
| Baseline | | 1,200.23 | 1,054.40 | 39.33 | 140.13 | 54.40 | 26.22 | 7.17 | 10.40 |
| April | | 441.56 | 623.12 | 11.24 | 33.46 | 27.19 | 25.18 | 7.20 | 6.38 |
| $50,000 - $60,000 | 1,794 | | | | | | | | |
| Baseline | | 1,247.15 | 1,151.32 | 51.36 | 175.07 | 52.08 | 25.84 | 7.18 | 10.72 |
| April | | 370.85 | 606.07 | 11.37 | 86.27 | 23.59 | 21.89 | 6.48 | 5.72 |
| $60,000 - $70,000 | 477 | | | | | | | | |
| Baseline | | 1,330.57 | 1,405.74 | 65.39 | 219.35 | 49.25 | 24.39 | 6.87 | 10.72 |
| April | | 345.54 | 454.35 | 10.94 | 30.24 | 22.74 | 21.35 | 6.47 | 8.60 |
| $70,000 - $80,000 | 249 | | | | | | | | |
| Baseline | | 1,362.35 | 1,292.92 | 78.90 | 204.47 | 49.69 | 24.34 | 6.76 | 10.55 |
| April | | 327.46 | 449.69 | 8.13 | 14.74 | 22.90 | 21.57 | 6.62 | 5.31 |
| $80,000 - $150,000 | 215 | | | | | | | | |
| Baseline | | 1,379.97 | 1,742.41 | 84.50 | 260.40 | 50.26 | 28.25 | 6.92 | 11.24 |
| April | | 356.94 | 551.27 | 15.85 | 65.83 | 22.57 | 24.44 | 6.68 | 5.89 |
| All | 11,398 | | | | | | | | |
| Baseline | | 1,221.46 | 1,039.68 | 40.35 | 140.04 | 56.42 | 29.32 | 7.16 | 4.80 |
| April | | 470.81 | 628.65 | 10.67 | 44.75 | 29.29 | 27.38 | 6.85 | 5.84 |

*Note. M* = mean; *SD* = standard deviation.

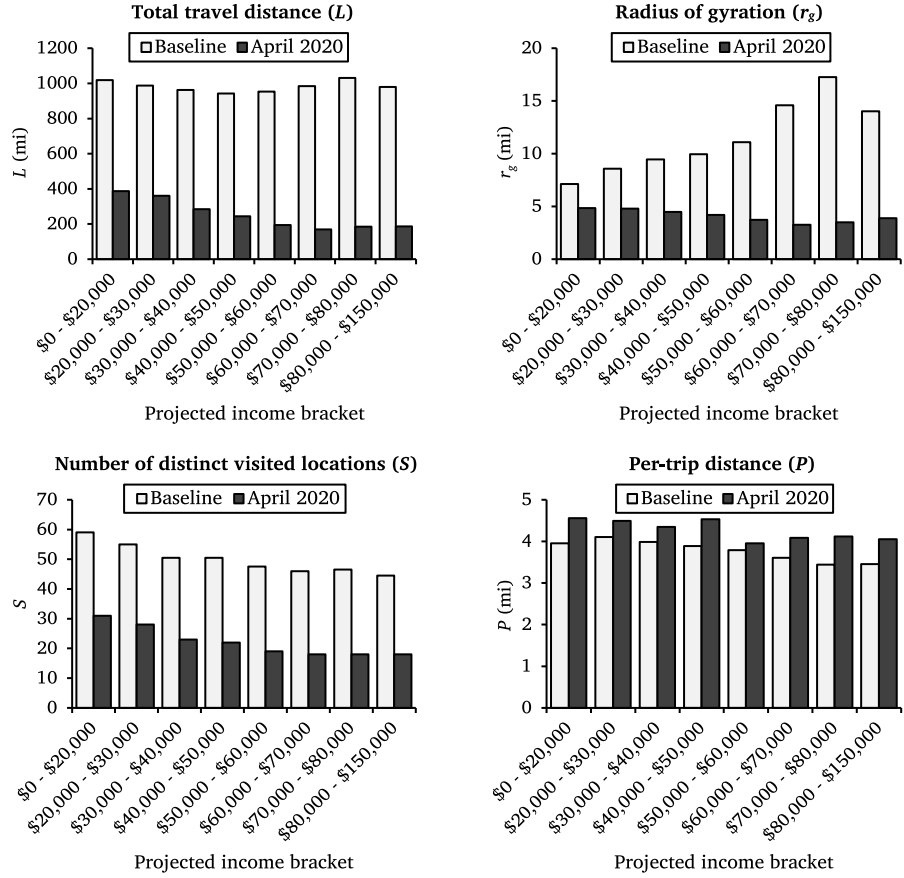

FIG. 3. Median total travel distance (top left), radius of gyration (top right), number of visited locations (bottom left), and per-trip distance (bottom right) by projected per-capita income bracket.

*Note.* See Table 1 for the sample size in each income bracket. See Fig. 4 through Fig. 7 for the interquartile range.

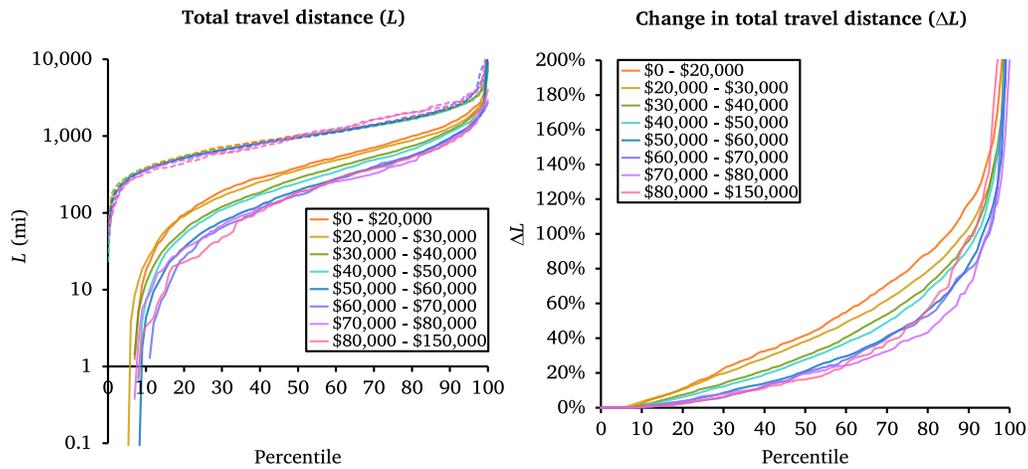

FIG. 4. Cumulative density functions of the total travel distance by projected per-capita income group.

*Note.* Parentheses indicate the sample size in each income bracket. Dashed lines indicate the baseline (average of January 2020 and February 2020) whereas solid lines represent April 2020.

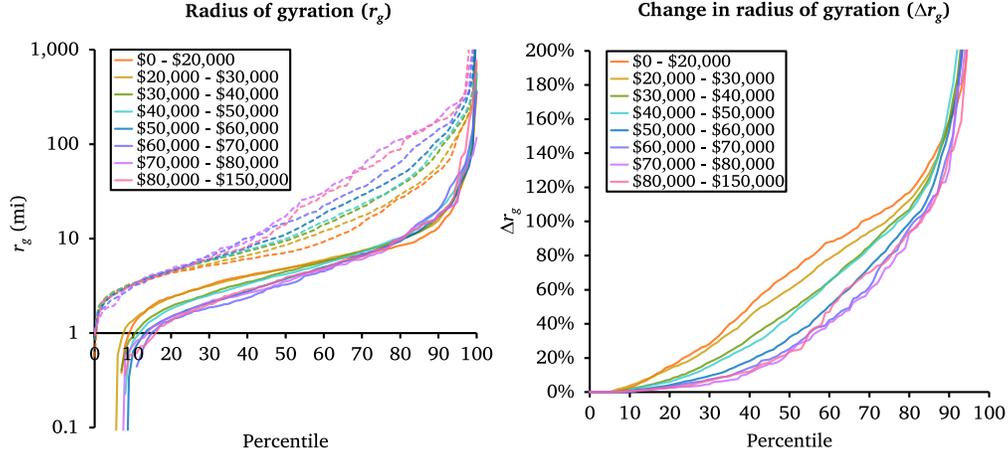

FIG. 5. Cumulative density functions of the radius of gyration by projected per-capita income group.

*Note.* Parentheses indicate the sample size in each income bracket. Dashed lines indicate the baseline (average of January 2020 and February 2020) whereas solid lines represent April 2020.

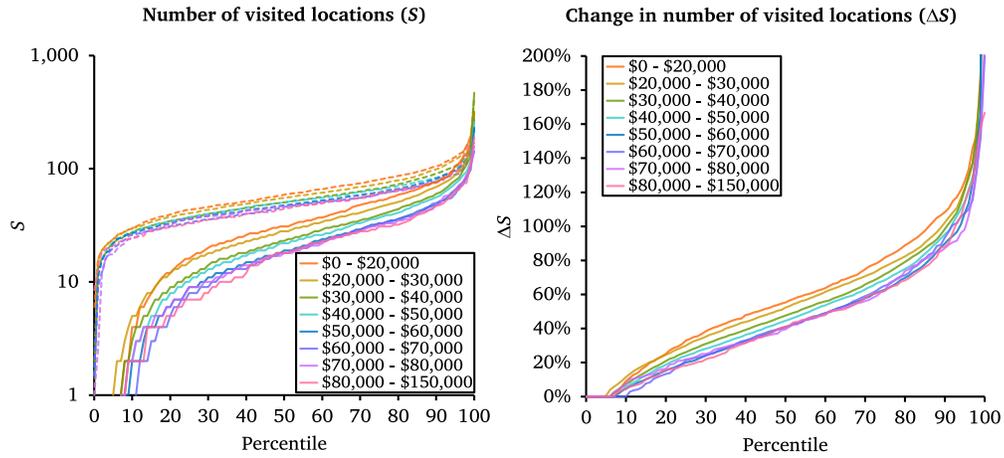

FIG. 6. Cumulative density functions of the number of distinct visited locations by projected per-capita income group.

*Note.* Parentheses indicate the sample size in each income bracket. Dashed lines indicate the baseline (average of January 2020 and February 2020) whereas solid lines represent April 2020.

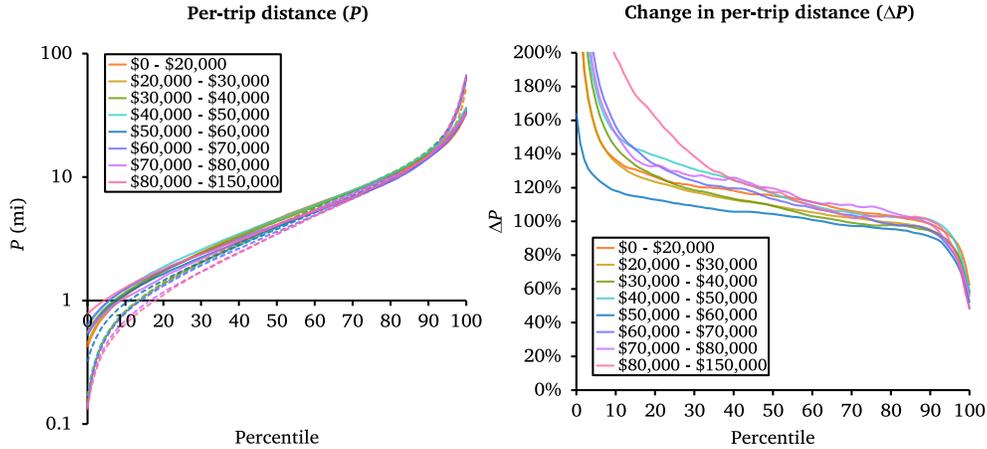

FIG. 7. Cumulative density functions of per-trip distance by projected per-capita income group.

*Note.* Parentheses indicate the sample size in each income bracket. Dashed lines indicate the baseline (average of January 2020 and February 2020) whereas solid lines represent April 2020.

## 5. DISCUSSION AND CONCLUSIONS

This study used pseudonymized smartphone location data to investigate the effects of the COVID-19 pandemic on individual human mobility from Greater Houston. This is part of an ongoing effort to understand mobility change during the COVID-19 pandemic in a greater detail. Our attempt was unique in that it implemented a longitudinal study design on (i) spatially disaggregated mobility data along with (ii) information from census tracts to capture mobility change (iii) in a single metropolitan statistical area with statewide executive orders to stay at home. This study investigated human mobility from four measures: total trip distance, radius of gyration, number of distinct visited locations, and per-trip distance.

### 5.1 Key Findings

Overall, our analyses revealed mobility adaptation disparity by income. The data indicated that human mobility in the Greater Houston area dropped significantly in April 2020 compared to the baseline (January and February 2020). We found that individuals from census tracts with a higher per-capita income had larger mobility reductions than people from tracts with lower per-capita income while the de facto stay-at-home orders were in effect.

In particular, the mobility disparity among income brackets was evident in total travel distances. Before the travel restriction, individuals tended to have similar travel distances regardless of estimated income brackets. However, higher-income groups saw a larger decrease in travel distances than lower-income groups in April. In April, the median travel distance was 387.23 mi for those whose estimated per-capita income was under $20,000 and 185.82 mi for those with estimated income was $80,000 and over.

Individuals from higher-income tracts had larger average radii of gyration before the pandemic; these individuals had slightly smaller radii of gyration during the executive orders than lower-income individuals, indicating that higher-income individuals reduced their radii of gyration more than lower-income individuals did. With the result of the travel distance alone, one might think the observed difference was largely due to the fact that tracts with higher income are located in the western center of Houston and thus enabled the residents to travel shorter distances during the executive orders. However, the smaller disparities in the radius of gyration and per-trip distance implies that people in higher-income tracts likely made fewer numbers of trips than those in lower-income tracts. The number of distinct visited locations followed a pattern similar to the total travel distance.

There was no clear disparity across income brackets in the per-trip distance. Interestingly, per-trip distance increased in April over the majority of percentiles regardless of the income brackets. In median, per-trip distances increased by less than a mile in all the estimated income groups. In general, trip distance distribution shifts towards zero when the radius of gyration becomes smaller (González et al., 2008). While the data do not tell the reason behind the observed increase, it is likely that there was a fundamental change in people's trip choice in the midst of the executive orders. Therefore, caution should be exercised when one tries to apply conventional travel choice models to data collected during travel restrictions. It is possible that people made fewer pass-by stops at attractions, such as restaurants and coffee shops, in April than they did in the baseline period. However, further research would be needed to identify the reasons behind the observed increase in per-trip distance.

Financial status was a likely factor of the compliance or feasibility of stay-at-home orders. Considering that an individual's "essential" fixed expenses to maintain the reasonable minimum standards of living (e.g., food, water, and electricity) do not differ vastly, it is intuitive that the disparity in mobility reduction seemed to be smaller among higher-income brackets than lower-income brackets (Fig. 3).

### 5.2 Implications of the Findings

The findings have meaningful social implications. It is likely that people in lower income brackets did not or could not reduce as much mobility as higher-income groups did during the executive orders in the midst of the COVID-19 pandemic. In total travel distance, radii of gyration, and the number of distinct visited locations, cumulative density functions revealed consistently lower mobility reductions for individuals in lower income brackets throughout most percentile ranges. While media and other articles have speculated that the disproportionality of COVID-19 impacts by different financial status, this is one of the first articles reporting the mobility adaptation disparity in the United States at the census tract level. If combined with further studies, the findings can be valuable for macroscopic and mesoscopic COVID-19 epidemiological model development and calibration as well as policy evaluations hereafter.

Our findings are consistent with Circella (2020), who reported difficulties in mobility adaptation among lower-income populations during the COVID-19 pandemic. Furthermore, the analyses supported what some nationwide studies in the United States (Jay et al. 2020; Weill et al., 2020) and abroad (Pepe et al., 2020) have also claimed: higher income is associated with larger mobility reductions. Since those preceding nationwide studies used aggregated location data to compare mobility across a nation, it is noteworthy that the disparity in mobility adaptation was observed among different income groups within a metropolitan statistical area where uniform executive orders were enforced. The results were in line with media speculations, which might explain a reason behind the disproportionate impacts of COVID-19: people with low income did not have a practical choice to stay at home. Because research indicates that racial residential segregation in Houston had already been associated with poor-self reported health before the COVID-19 outbreak (Anderson and Oncken, 2020), it was likely that the pandemic had amplified inequity in the society.

### 5.3 Limitations and Future Research Directions

Our research had some limitations. One limitation of the present study, as well as existing literature, was that the mobile travel data may exhibit a systematic error. In other words, pseudonymized individuals may not represent each demographic group without bias. The authors still maintain reasonable confidence due to the fact that the sample sizes are all fairly large across the groups. In future studies, this limitation can be addressed if sample data can be stratified based on demographics information directly tied to mobility data.

Another limitation was that the authors did not overlay any other socioeconomic variables (e.g., gender or household size) associated with each census tract. This was mainly because we preferred to avoid stacking ecological fallacies. However, it has been known that age is a major covariate of income as income tends to show an inverted U-shape with age (Sturman, 2003). Although it is difficult to eliminate the effects of confounders completely, it would be ideal to control the variable if the data availability and a research design allows. Researchers would get results closer to the "ground truth" if the true demographics of pseudonymized mobility data were available. For example, essential businesses, such as grocery stores and gas stations, remained open during the executive orders, but there is little research that took such factors into account because such mobility data are rarely available to the public. When it comes to economic status, disposable income, current assets, and net assets could be better indicators of financial leeway than pre-tax income if such variables are available in future research.

While this study showed an apparent disparity in travelers' mobility patterns during the pandemic, this study could not distinguish the trip purposes or the movements of specific types of workers. To reveal the reasons behind mobility pattern change, more qualitative research would be worth conducting. For instance, surveys about typical travel purposes before and during the pandemic may reveal the reasons behind the shift in travelers' mobility (e.g., increased per-trip distance) by controlling covariance.

This study demonstrated the potential of the use of spatially disaggregated data on mobility research. Similar regional research in different states and countries would further reveal comparative impacts and generalizability of the associations between financial status and mobility disparity during travel restrictions. In the near future, it would be beneficial to conduct meta-analyses to integrate insights about mobility changes related to the COVID-19 pandemic.

### ACKNOWLEDGMENTS AND FUNDING RESOURCES

We acknowledge SAFE2SAVE, LLC for providing data under specific terms and conditions.

This research did not receive any specific grant from funding agencies in the public, commercial, or not-for-profit sectors.